# Machine learning-based analysis of ECG and PCG signals for rheumatic heart disease detection: A scoping review (2015–2025)

Damilare Emmanuel Olatunji [*,1], Julius Dona Zannu, Carine Pierrette Mukamakuza [*], Godbright Nixon Uiso, Chol Buol, John Bosco Thuo, Nchofon Tagha Ghogomu, Mona Mamoun Mubarak Aman, Evelyne Umubyeyi

*College of Engineering, Carnegie Mellon University Africa, Kigali, Rwanda*

ABSTRACT

AI-powered stethoscopes offer a promising alternative for screening rheumatic heart disease (RHD), particularly in regions with limited diagnostic infrastructure. Early detection is vital, yet echocardiography, the gold standard tool, remains largely inaccessible in low-resource settings due to cost and workforce constraints. This review systematically examines machine learning (ML) applications from 2015 to 2025 that analyze electrocardiogram (ECG) and phonocardiogram (PCG) data to support accessible, scalable screening of all RHD variants in relation to the World Heart Federation's "25 by 25" goal to reduce RHD mortality. Using PRISMA-ScR guidelines, 37 peer-reviewed studies were selected from PubMed, IEEE Xplore, Scopus, and Embase. Convolutional neural networks (CNNs) dominate recent efforts, achieving a median accuracy of 97.75 %, F1-score of 0.95, and AUROC of 0.89. However, challenges remain: 73 % of studies used single-center datasets, 81.1 % relied on private data, only 10.8 % were externally validated, and none assessed cost-effectiveness. Although 45.9 % originated from endemic regions, few addressed demographic diversity or implementation feasibility. These gaps underscore the disconnect between model performance and clinical readiness. Bridging this divide requires standardized benchmark datasets, prospective trials in endemic areas, and broader validation. If these issues are addressed, AI-augmented auscultation could transform cardiovascular diagnostics in underserved populations, thereby aiding early detection. This review also offers practical recommendations for building accessible ML-based RHD screening tools, aiming to close the diagnostic gap in low-resource settings where conventional auscultation may miss up to 90 % of cases and echocardiography remains out of reach.

## 1. Introduction

Rheumatic heart disease (RHD), a challenging condition caused by group A streptococcus, remains a significant global health challenge with devastating consequences [1–3]. In 2021, 373,000 (326,000–446, 000) died due to RHD globally [4]. Current epidemiological data indicate that RHD affects around 40.5 - 50 million people globally, and results in approximately 360,000 deaths each year [5–7]. Equally, the burden of this disease falls disproportionately on low and middle-income countries (LMICs) - Fig. 1 [8], where it accounts for up to 1.5 % of all cardiovascular disease-related mortality [9]. This disparity highlights treatment gaps across regions. In response, the World Heart Federation (WHF) and WHO aim to reduce RHD-related premature mortality by 25 % by 2025, emphasizing early detection as a key strategy [9,10].

However, this focus on early detection unveils a paradox in managing RHD. Auscultation proficiency (Fig. 2) declining due to limitations in the human auditory system and clinicians' skills [11,12]. The diagnostic gold standard, echocardiography, which could detect 3–10 times more RHD cases than clinical auscultation, remains inaccessible in low and low-medium-income countries. This inaccessibility stems from high costs and equipment shortages, leaving most individuals undiagnosed until complications arise [13–16]. This diagnostic gap has sparked interest in developing alternative, accessible screening methods for high-prevalence, resource-limited settings.

Electrocardiography (ECG) and Phonocardiography (PCG) have emerged as promising candidates for this diagnostic challenge, offering advantages over echocardiography in resource-limited settings. ECG records the heart's electrical activity and detects conduction abnormalities linked to RHD progression [17,18]. PCG captures acoustic cardiac signals, including murmurs from valvular pathologies central to RHD pathophysiology [19,20] - Fig. 3. Their simplicity, non-invasiveness, and lower resource needs make them appealing for screening. Recent innovations and a study by [21] suggest these modalities could serve as effective screening filters by general physicians or primary healthcare clinicians in low-resource communities, reducing echocardiography needs while maintaining acceptable disease detection sensitivity [22]. After acquisition, the recorded signal is processed either on-device or transferred to a connected edge device where the ML model performs automated analysis. The model then outputs a simple diagnostic result (e.g., normal vs abnormal), supporting rapid triage and referral where necessary [23–25].

* Corresponding authors.
*E-mail addresses:* dolatunj@andrew.cmu.edu (D.E. Olatunji), cmukamak@andrew.cmu.edu (C.P. Mukamakuza).

[1] Present Address: Regional ICT Center of Excellence Bldg Plot No A8, Kigali Special Economic Zone Phase II, Kigali, Rwanda.

Within cardiac diagnostics specifically, ML approaches have demonstrated remarkable efficacy in analyzing ECG and PCG signals across multiple rheumatic heart disease pathologies. For instance, [25] reported a lightweight hybrid deep learning system (CNN and LSTM) for cardiac valvular disease classification of five heart valvular conditions, namely normal, aortic stenosis, mitral regurgitation, mitral stenosis and mitral valve prolapse, achieving a 93.76 % accuracy, 85.59 % F1-score and AUC of 0.9505. While [26] developed supervised machine learning models for RHD classification in Ethiopia with SVM achieving a surpassing accuracy of 96 % over KNN with 88 %, Logistic regression - 92 % and Random Forest - 90 %. These advances suggest clinical value proposition and compelling potential for RHD detection as characteristic valvular abnormalities produce distinctive electrical and acoustic signatures potentially identifiable, potentially averting thousands of disability-adjusted life years through earlier intervention.

Remarkably, despite the evident potential of ML-based ECG and PCG analysis for RHD detection, the research landscape remains fragmented and inadequately characterized [27]. Individual studies have reported encouraging results using various algorithms and signal processing techniques, but there has been no comprehensive review that systematically organize evidence, maps out state-of-the-art approaches, catalog the different methodologies, analyzes patterns across literature, and identifies where research gaps exist. As such, this study aim to observe different RHD related studies, dataset characteristics, methodologies, and performance metrics to enable cross-study comparison.

Guided by the PRISMA-ScR framework, our scoping review aims to:

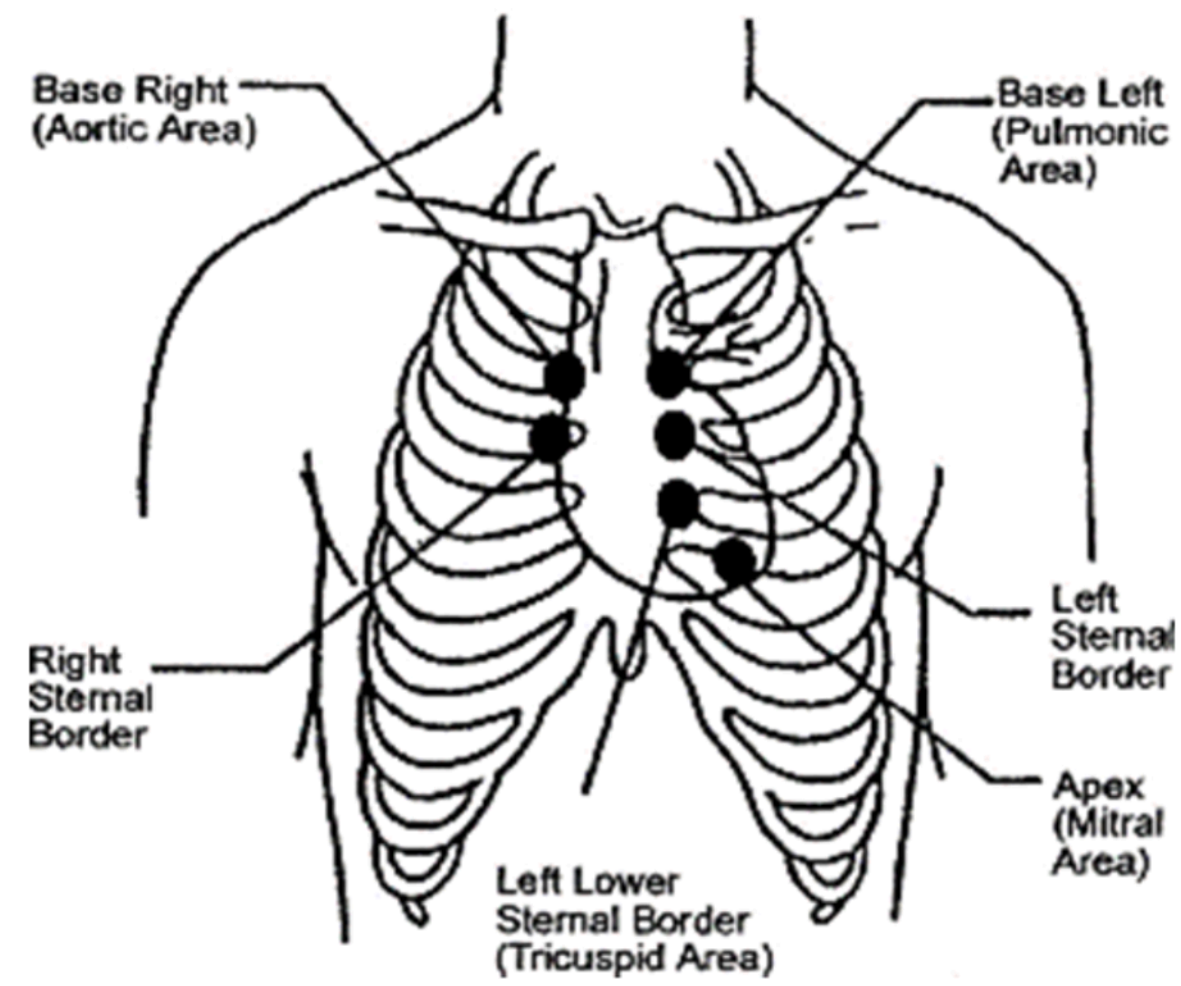

**Fig. 2.** Anterior chest wall auscultation points corresponding to the four cardiac valve areas. These landmarks aid clinicians in detecting murmurs and abnormal heart sounds linked to valvular diseases such as rheumatic heart disease.

(1) comprehensively map machine learning approaches using ECG and PCG signals for RHD detection published between 2015 and 2025 (2) categorize the prevalent ML techniques, feature extraction methods, and signal processing approaches (3) characterize the validation methodologies, performance metrics, and dataset characteristics reported across literatures and (4) identify research gaps, and opportunities for clinical implementation. Our selected timeframe of 2015–2025 aligns with the World Heart Federation's "25 by 25" goal to reduce RHD mortality. By incorporating recent research, we aim to provide insights into current trends and patterns for researchers, technology developers, and clinicians, emphasizing both strengths and weaknesses to guide future studies.

## 2. Methods

This scoping review followed PRISMA-ScR guidelines [28] and was registered with the Open Science Framework (OSF). It focused on original research articles and conference proceedings related to AI applications in analyzing ECG and PCG signals for diagnosing rheumatic

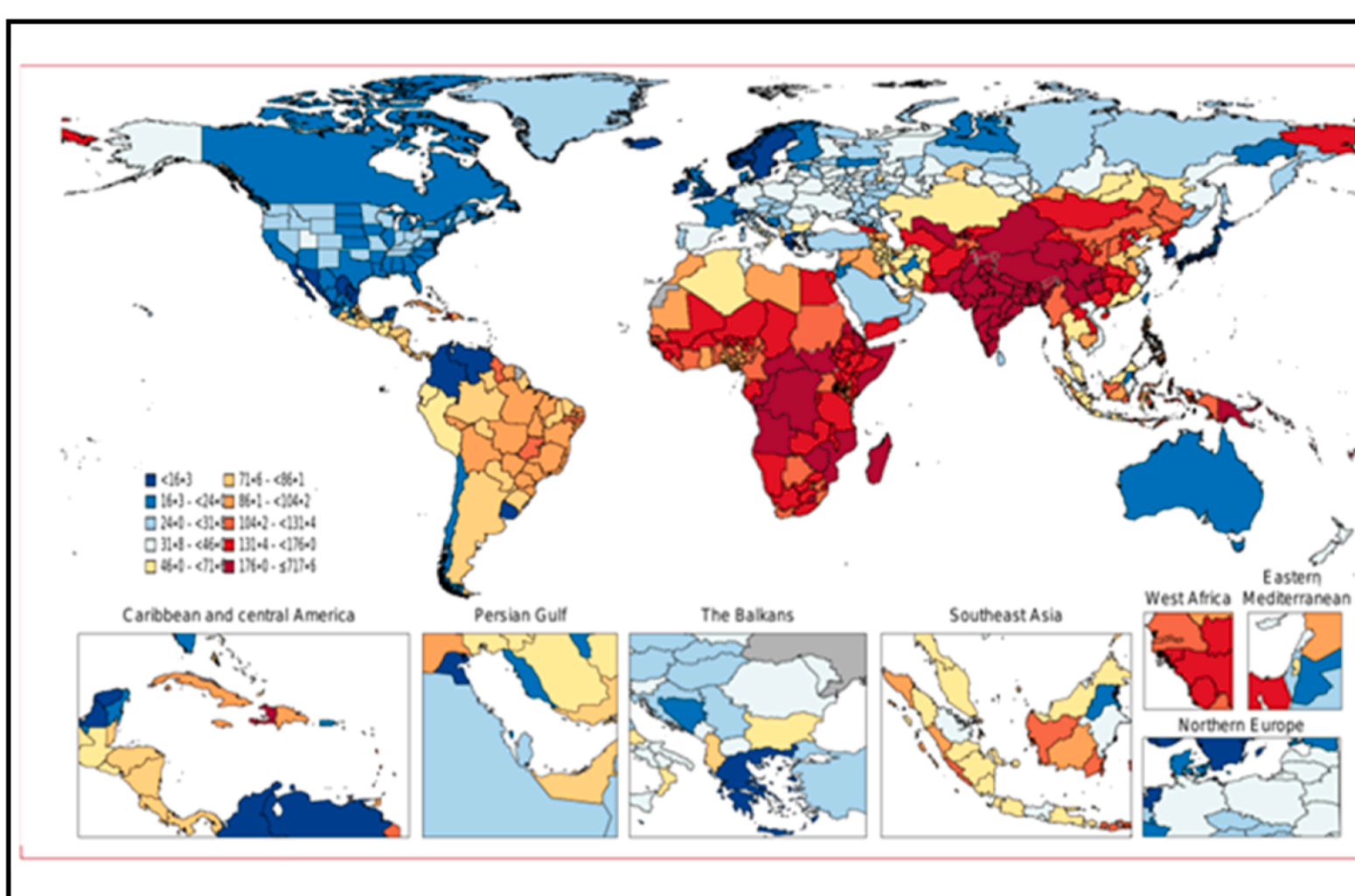

The gradient color scale represents different DALY rates:

- **Dark blue (lowest rates)**: <163 DALYs per 100,000 people
- **Light blue (low rates)**: 163 - 243
- **Yellow (moderate rates)**: 318 - 486
- **Orange to Red (high rates)**: 861 - 1,024 or more
- **Dark red (highest burden)**: 1,760 - 5,717

**Key observations from the Map**

Sub-Saharan Africa has the highest disease burden due to infectious diseases, malnutrition, and limited healthcare access. South Asia, Central Asia, and Eastern Europe also experience high DALY rates, primarily driven by cardiovascular diseases, respiratory infections, and non-communicable diseases. The Middle East and West Africa show high disease burdens, influenced by conflict-related injuries, maternal health issues, and infectious diseases.

South America, Southeast Asia, and parts of the Middle East fall into the moderate disease burden category, indicating a combination of communicable and non-communicable diseases. These regions have been improving but still have insufficient healthcare systems.

North America, Australia, Western Europe, and parts of East Asia, including Japan, South Korea, and China, have the lowest disease burden. These regions benefit from strong healthcare systems, better living conditions, and a lower prevalence of infectious diseases.

**Fig. 1.** Global age-standardized Disability-Adjusted Life Year (DALY) rates per 100,000 population for rheumatic heart disease (RHD) in 2021, both sexes combined. The map highlights the disproportionate burden of RHD across low- and middle-income countries (LMICs), particularly in sub-Saharan Africa, South Asia, and parts of Oceania and the Middle East.

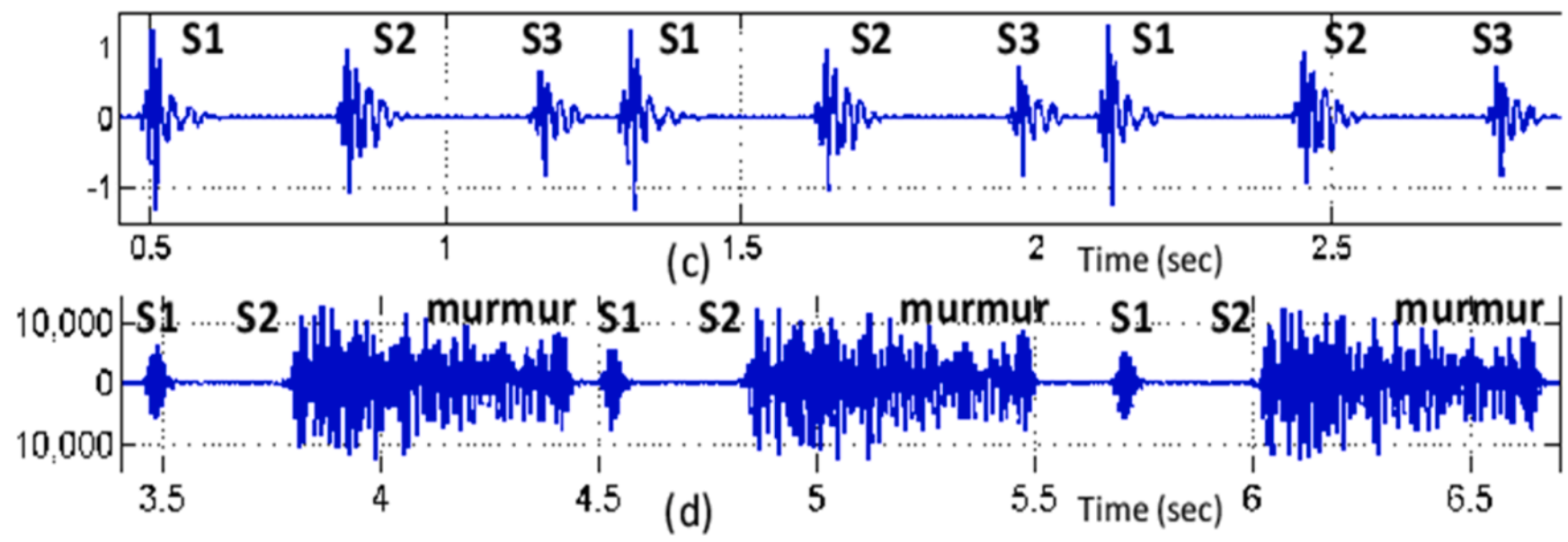

**Fig. 3.** Phonocardiogram (PCG) recordings showing normal heart sounds (S1, S2, S3) and pathological murmurs characteristic of valvular diseases. Murmurs, as seen in the lower panel, are prolonged acoustic events occurring between normal heart sounds and are key indicators of rheumatic heart disease.

heart disease and its variants from January 2015 to March 2025. The databases searched included PubMed, Scopus, IEEE Xplore, and Embase. The search strategy for PubMed is presented in the appendix (Table I). Two independent reviewers screened the titles and abstracts, categorizing them as "include," "exclude," or "uncertain," with agreement assessed using Cohen's Kappa coefficient. Selected articles underwent another full-text review using a standardized form (Fig. 4).

## 3. Results

This section presents the synthesis of studies identified through our systematic search. The PRISMA flow diagram documents the selection process, including the identified, screened, and included records. Findings are organized thematically, highlighting key patterns across the studies. Descriptive statistics of study characteristics (methodology, geography, sample sizes) precede the analysis of primary outcomes. Gaps in current knowledge and areas of consensus/disagreement are identified. Findings are presented neutrally, with interpretations reserved for the discussion section.

A. Descriptive statistics of study characteristics (methodology, geography, sample size, research outcome, and limitations identified).
1. **Rheumatic Heart Disease (RHD) Detection:** Table 3 shows the studies, primarily focusing on detecting RHD using either PCG or ECG-based features.

2. **Aortic Stenosis (AS) Detection:** The studies in Table 4 utilize either ECG or acoustic (stethoscope or audio) signals to detect aortic stenosis.

3. **Aortic regurgitation (AR) Detection:** The studies in Table 5 leveraged either ECG or PCG signals to detect aortic regurgitation.

4. **Mitral Regurgitation (MR) Detection:** Table 6 studies focus on detecting mitral regurgitation using acoustic and ECG signals.

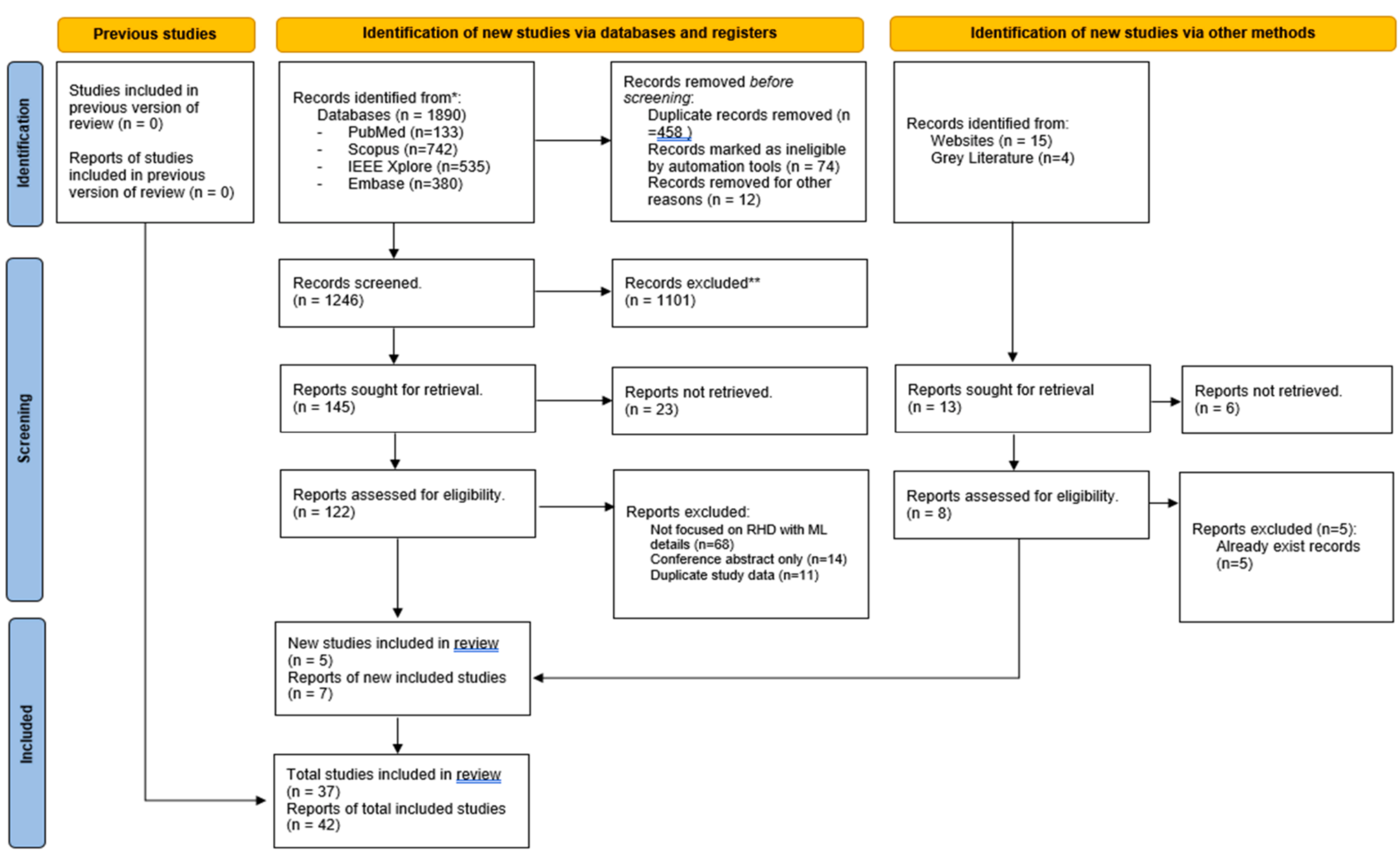

**Fig. 4.** PRISMA 2020 flow diagram illustrating the literature selection process for studies applying machine learning to cardiac signal analysis (ECG/PCG) in rheumatic heart disease detection between 2015 and 2025.

**Table 3**
Characteristics of machine learning studies for rheumatic heart disease detection.

| Ref | Year | Model | Preprocessing technique | Country | Sample size, data type, and access condition | Metric | Limitations |
|---|---|---|---|---|---|---|---|
| [11] | 2022 | Cubic SVM and Fine-KNN | Downsampled from 44.1 kHz to 2 kHz, split into 5-second windows, and extracted 26 features (time, frequency, and MFCC) using a 30 % holdout validation approach. | Ethiopia | 170 samples (124 RHD cases; 46 normal cases), PCG, Private | Accuracy: 97.1 %<br>Sensitivity: 98 %<br>Specificity:95.3 %<br>Precision: 97.6 % | Dataset had a 3:1 class imbalance. RHD severity levels were not considered. No comparison made with clinician accuracy. Age difference: RHD 22.9 ± 8.9 years; Normal 14.4 ± 10.5 years. |
| [29] | 2020 | CNN | Downsample from 44.1 kHz to 2 kHz, segment data into 1.2 s, transform 1D heart sound to 2D Log Mel Spectrogram, and normalize data. | Ethiopia | 170 subjects (124-RHD, 46 - Normal), PCG, Private | Accuracy:96.7 %,<br>Sensitivity: 95.2 %,<br>Specificity:98.2 % | Dataset imbalance, single-center collection, No external validation, minimal comparative analysis, and demographic disparities. |
| [30] | 2022 | Logistic regression, Random Forest, Deep Neural Network | Used multivariate outlier detection, 85/15 for training/testing, applied winsorization, conducted data normalization, and performed 10-fold cross-validation. | Pakistan | 561 subjects, PCG, Private | ROC: 0.901<br>(95 % CI: 0.818–0.983)<br>Sensitivity: 85.1 %<br>Specificity: 70.6 % | Single-center design, class imbalance (RHD MR: 75.94 % vs. RHD MS: 24.06 %), no external validation, and insufficient guidance on the developed model. |
| [31] | 2015 | SVM classifier and Hidden Semi-Markov Models (HMMs) | Sample entropy, kurtosis, and SVD for signal quality, applied Hamming windowing (50 % overlap), extracted features via Hilbert/wavelet envelopes and PSD, and bandpass-filtered (<200 Hz). | South Africa | 150 samples, PCG, Private | Signal quality classifier accuracy >90 %; F1 score for segmentation: 93.5 % | No classification of heart sounds for RHD. No clinical validation of the system's diagnostic capability for RHD. |
| [32] | | KNN, SVM, Logistic Regression, and Random Forest | Data cleaning, handling missing values, and feature selection from clinical and echocardiographic data | Ethiopia | 244 patient records, ECG, Private | Accuracy of models: KNN (88 %), RF (90 %), LR (92 %), SVM (96 %) | Limited data size, no external validation, and insufficient comparison with established clinical protocols. |
| [33] | 2023 | Not provided | Not provided | Sudan | 115 patients, clinical/ECG data, Private | Atrial fibrillation (18.3 %), and sinus rhythm (81.7 %), correlate with mitral stenosis | Single-center study, convenience sampling, limited demographic diversity. |
| [12] | 2022 | SVM with RBF kernel | Bandpass filtering (20Hz-1 kHz), downsampling to 2 kHz, z-score normalization, 30-second segment extraction | Ethiopia | 251 subjects (124 PwRHD, 127 HC), PCG data, Private (partially open via PhysioNet) | Stratified 10-fold CV:<br>F1=96.0 ± 0.9 %,<br>Recall=95.8 ± 1.5 %,<br>Precision=96.2 ± 0.6 %,<br>Specificity=96.0 ± 0.6 % | RHD severity not considered. Significant age differences (RHD mean age 22.9 ± 8.9 years) and (HC mean age 14.4 ± 9.4 years) |
| [34] | 2023 | ResNet-34 (1D CNN), CNN-LSTM hybrid model | Eighth-order Butterworth band-stop filter (20–200 Hz), downsampling to 2000 Hz, edge noise removal by cropping | India | PhysioNet 2016 (3126), CirCor DigiScope (5272), 5-class set (1000); Public | Specificity - 95.00 %<br>Sensitivity - 98.75 %<br>Accuracy - 98.00 % | High false positives from artifacts, poor performance in real-world data, underrepresentation of diastolic murmurs, inability to differentiate severity levels. |

**5. Mitral Stenosis (MS) Detection:** The studies in Table 7 focus on using both PCG and ECG signals, reporting robust classification metrics for mitral stenosis.

6. **Combined Valvular Heart Disease (VHD) Screening and Multi-Condition Detection:** Table 8 contains additional valvular conditions combined.

## 4. Discussion

a. Evolution of Machine Learning Approaches in RHD Detection

From classical algorithms to deep learning approaches, our review of 37 studies (2015–2025) shows a clear evolution in machine learning for RHD detection (Fig. 7). Early studies (2015–2019) predominantly employed SVMs and other traditional methods with manually engineered features, as exemplified by [37]. A pivotal shift occurred in 2020 toward CNNs, which became the dominant architecture by 2022–2025, appearing in 19 studies (51.4 %). Concurrently, hybrid models emerged from 2021 onward, with studies such as [21] and [55] combining CNN architectures with LSTM or ensemble methods to achieve accuracy improvements of 3–5 % over single-algorithm approaches. Despite this evolution towards deep learning, traditional algorithms remain prevalent, as indicated by recent studies like [50] in 2024, which showcase the effectiveness of SVM and Random Forest for specific RHD variants. This shows a practical acknowledgment that simpler models are valuable for interpretability. The evolution in modeling includes a shift from expert-defined features to end-to-end learning.

**b. Signal Processing and Feature Engineering Practices**

Fig. 8 reveals patterns in signal processing, with bandpass filtering as the primary technique (54.1 %), applied in the 20–250 Hz range to isolate cardiac sounds from noise. Downsampling (48.6 %) and normalization (43.2 %) were also common, establishing standards for preprocessing cardiac signals. This consistency indicates a consensus on optimal signal characteristics for RHD-related pattern recognition.

A clear divide emerged in feature extraction between traditional and deep learning methods. Traditional studies focused on explicit feature engineering, particularly MFCC extraction (35.1 %) and segmentation (32.4 %). In contrast, deep learning methods like CNNs often used transformed signal representations, such as spectrograms (27.0 %) or wavelets (21.6 %), for automatic feature discovery. This divide shows the trade-off between interpretability and performance-engineered features offer clearer clinical insight but may have lower discriminative power than automatically extracted features.

c. Performance Characteristics and Methodological Limitations

Tables 1–6 summarize technical performance metrics from 37

**Table 4**
Characteristics of machine learning studies for detecting aortic stenosis, a form of rheumatic heart disease.

| Ref | Year | Model | Preprocessing technique | Country | Sample size, data type, and access condition | Metrics | Limitation |
|---|---|---|---|---|---|---|---|
| [35] | 2020 | Multilayer perceptron (MLP) and CNN. | Hamming window, 50 % overlap, Hilbert envelopes, wavelet envelopes, power spectral density, normalization, bandpass filtering. | South Korea | 29, 859 samples, ECG, Private | AU-ROC: 0.884 (internal), 0.861 (external) | No comparison with other algorithms' performance and cardiologists' interpretations. Limited insight into the T-wave role in detecting aortic stenosis. |
| [36] | 2022 | CNN | Background noise elimination, recording trimming (beginning and ending) placement and removal of click noises. | Israel | 100 samples, PCG, Private | Sensitivity 90 %, Specificity 84 % | No comparison with clinicians' auscultation proficiency. Exclusions of specific conditions and limited patient data on mitral regurgitation. |
| [37] | 2022 | CNN | Recorded at 40 kHz, segmented into 5-second clips, MFCCs extraction, and batch normalization with dropout ($p$ = 0.2) | USA | 240 patients, PCG, Private | Sensitivity - 0.90 (0.81–0.99) Specificity - 1 F1-score - 0.95 (0.89–1.0) | The algorithm was trained to detect aortic stenosis, limiting its applicability in real-world situations where multiple valve pathologies may exist. |
| [38] | 2015 | Hybrid model - HMM with SVM | Calculate spectral energy using temporal sliding windows over discriminative frequency bands, followed by quantification using Mahalanobis distance. | Sweden | 50 patients; PCG with synchronous ECG; Private | Average accuracy: 81.7 % (95 % CI), Sensitivity: 79.3 % (95 % CI), Specificity: 82.9 % (95 % CI). | The dataset includes only 50 patients, limiting the representation of aortic stenosis. Mild and moderate cases were neglected, and no validation on the external dataset. |
| [39] | 2023 | CNN - Transfer learning and XGBoost | Median pass filtering, scaling to millivolts, normalizing, and noise addition with random Gaussian fluctuations in different frequency ranges. | USA | 75,901 ECG-TTE pairs, from 35, 992 unique patients, ECG, private | AUROC of 0.829 (95 % CI) for detecting moderate/ severe AS and 0.846 (95 %) for severe AS. | Low specificity of 58.7 % (many false positives). Single-center dataset, lack external validation and prospective analysis for community screening. |
| [40] | 2019 | SVM | Spectral noise subtraction, fourth-order Butterworth band-pass filtering (25–140 Hz), automated heartbeat segmentation, systolic interval extraction, Hilbert transform, and low-pass filtering. | USA | 96 Subjects (12 AS, 84 - No AS), PCG synchronized with ECG, Private | Sensitivity - 92 % Specificity - 95 %, ROC curve (AUC) - 0.94 for amplitude feature and 0.87 for spectral feature | The algorithm relies solely on two features-amplitude and frequency center of mass-and lacks validation on an external cohort from diverse clinical settings. |
| [41] | 2021 | CNN - A DenseNet with 63 layers (classification inclusive) | ECG signal upsampling from 250 Hz to 500 Hz using the 'Resample' function, ECG matrix preparation (12 × 5000 dimensions), zero-padding, and normalization | USA | 258,607 patients with ECG-TTE pairs; ECG, Private | With age and sex included: AUC = 0.87. For patients without hypertension: AUC = 0.90 (sensitivity = 75 %, specificity = 88 %). | No systematic assessment of cardiac murmurs, hindering understanding of AI-ECG performance in affected patients. No external dataset for validation. |
| [42] | 2024 | CNN with depthwise and residual connections. | Bandpass filtering, window slicing with a 1 s sliding window, value embedding using 1D circular padding, and position embedding using sine and cosine functions. | China | Approx 80 subjects (38AS, 40 Norm), PPG, Private | Accuracy (91 % ± 0.03), Sensitivity (93 % ± 0.05), Specificity (89 % ± 0.01), F1-score (90 % ± 0.03), AUC (91 % ± 0.03). | Relatively small dataset. Exclusion of patients with comorbid arrhythmias and other valvular diseases that affect hemodynamics. |

**Table 5**
Characteristics of machine learning studies for detecting aortic regurgitation - RHD variant.

| Ref | Year | Model | Preprocessing Technique | Country | Sample size, data type, and access condition | Metrics | Limitation |
|---|---|---|---|---|---|---|---|
| [43] | 2024 | CNN - ResNet | Baseline wander removal, low-pass filtering, standardization, segmentation of lead II into 4 segments of 2.5 s, and alignment of leads | Japan & Taiwan | 573 patients, 1,45,712-lead ECGs, private | AUROC: LVESDi >20 mm/m$^2$ (0.85), LVESDi >30 mm/m$^2$ / LVESVi >45 ml/m$^2$ (0.84), LVEF <40 % (0.83) | No validation on external cohorts and only focuses on moderate-severe or severe AR, limiting applicability to all AR severity levels. |
| [44] | 2022 | 2D-CNN +FC-DNN | Raw ECG data (5000 × 12 matrix, 500 Hz sampling), augmentation via stride extraction, z-score normalization | South Korea | 29,859 ECG-echocardiography pairs (412 AR cases), private | Multi-input model: AUROC=0.802 (95 % CI), Sensitivity=53.5 %, Specificity=82.8 %, PPV=5.0 %, NPV=99.1 %, 2D-CNN alone: AUROC=0.734 ($p < 0.001$) | Lacked validation with external data. Failed to assess clinical performance and did not provide information on patients' heart failure status or atrial fibrillation (AR) etiology. |

studies, showing a median accuracy of 97.75 % (IQR: 88.4–97.1 %) and AUROCs often over 0.85. Notable findings include [56] with 99.32 % accuracy for VHD detection, [61] at 99.90 % accuracy using Vision Transformers, and [21] with 99.87 % multiclass classification accuracy. Nonetheless, these studies require careful assessment due to some identified methodological limitations.

Sample size inadequacies represent a primary concern, with a median cohort of only 244 subjects (IQR: 170-617). Studies like [37] ($n$ = 50), [35] ($n$ = 100), and [58] ($n$ = 15) highlight this limitation, while class imbalances were prevalent in studies such as [10] and [28] where RHD cases outnumbered controls by nearly 3:1. Validation deficiencies were equally problematic, with only 4 studies (10.8 %) reporting external validation on independent cohorts. Studies [34,46], and [42] demonstrated performance degradation during external validation (e.g., AUROCs dropping from 0.89→0.86 for AS and 0.88→0.81 for MR in [52]), emphasizing this critical weakness.

**Table 6**
Characteristics of machine learning studies for detecting Mitral Regurgitation - RHD Variant.

| Ref | Year | Model | Preprocessing Technique | Country | Sample size, data type, and access condition | Metrics | Limitation |
|---|---|---|---|---|---|---|---|
| [45] | 2024 | Clique block-based DNN | Two-stage noise cancellation, 2 s sliding window segmentation | China | 823 sample size, PCG, Private 4 severity classes (none, mild, moderate, severe) | Sensitivity = 85.6 % Specificity = 84.4 % | Limited dataset size. No ECG-based segmentation |
| [46] | 2021 | KNNs, Adaboost, and SVMs | Standardized signals to 3 s segments, noise removal, bandpass filtering (20–250 Hz), and extracted time-domain (peak amplitude, duration, ZCR) and frequency-domain (7 MFCCs) features | India | Open-source PhysioNet CinC Challenge and PASCAL heart sound dataset. | For SVM: Accuracy ≈92.77 %, Sensitivity ≈85.48 %, Specificity ≈94.22 %, F1-score ≈86.40 % (Adaboost: 100 % across metrics) | Lacks validation in real patient settings. While it mentions using Zero-Crossing Rate (ZCR), MFCCs, and peak amplitude features, it fails to justify their selection over other options. |
| [47] | 2020 | CNN | Recorded 12-lead ECG (500 Hz, 8 s) with noise filtering and normalization. Used raw ECG (60,000 features) in a CNN with residual blocks, batch normalization, and dropout (0.2), and visualized decisions with Grad-CAM maps. | South Korea | 56, 670 ECGs from 24, 202 patients, private | 12-lead ECG: Internal AUROC 0.816, External AUROC 0.877, single-lead ECG: AUROC 0.758 (internal), 0.850 (external) | The study exhibited class imbalance, with MR prevalence at 3.9 % in the external validation dataset versus 25.96 % internally. It also failed to compare the AI algorithm's performance with cardiologists and lacked prospective validation in real-world screening settings. |
| [48] | 2024 | CNN - ResNet18 | Upsampled to 500 Hz before analysis. | USA | 4019 patients, ECG, Private | AUC for identifying diastolic dysfunction grades > (1,2, 3) - 0.847, 0.911, 0.943 | The AI-ECG model's interpretability is limited, as the specific electrocardiographic features influencing classification are not fully clear, although saliency maps offer some insight. |

**Table 7**
Characteristics of machine learning studies for detecting Mitral Stenosis - RHD Variant.

| Ref | Year | Model | Preprocessing Technique | Country | Sample size, data type, and access condition | Metrics | Limitation |
|---|---|---|---|---|---|---|---|
| [49] | 2020 | Decision Tree Learning (DTL) model | Reduced to 500 samples/s with 5 μV resolution using Philips ECGVue; automated ECG feature extraction | China | 59 ECGs from 59 mitral stenosis patients, private | Accuracy = 0.84, Precision = 0.84, Recall = 0.83, F-measure = 0.84) | Single-center study with a small sample size and retrospective analyses |
| [50] | 2021 | Univariate and multivariate logistic regression | None - study focused on manual measurement of P-wave parameters from standard 12-lead ECGs. | China | 124 subjects (62 MS patients and 62 healthy controls); ECG, Private | Maximum P-wave duration (OR: 1.221, 95 % CI: 1.126–1.324) and P-wave dispersion (OR: 1.164, 95 % CI: 1.094–1.238) | Insufficient dataset, the sample size is inadequate to stratify findings across mild, moderate, and severe MS subgroups. Single-center dataset. |
| [51] | 2024 | SVM, Random Forest, KNN, and Decision Tree | Down-sampling of PCG signals to 1000 Hz and de-noising using Bior-4.4 wavelet to remove high-frequency components. | India | 5002 PCG signals, combined (Public and Private) | Accuracy (RF = 98 %, SVM = 97.6 %, DT = 97.2 % and KNN = 96 %). | No temporal validation to assess the model's stability over time and different recording conditions. |

1. Demographic bias emerged as another significant limitation, with 18 studies (48.6 %) reporting substantial age and gender differences between test groups. For example, [10] and [11] noted age disparities between RHD patients (mean 22.9 ± 8.9 years) and controls (14.4 ± 10.5 years), potentially introducing confounding factors. Additionally, 26 studies (70.3 %) had limited severity stratification, treating RHD detection as a binary issue instead of addressing critical disease severity gradations. Only [44] and [56] attempted multi-class severity stratification, vital for management decisions.
2. The lack of comparison with clinical standards is a significant gap, with only 5 studies (13.5 %) comparing algorithm performance to expert clinician auscultation. Additionally, only 8 studies (21.6 %) calibrated metrics to realistic disease prevalence, with study [11] showing F1-scores drop from 96.0 % to 72.2 % at 5 % prevalence. Another study [57] highlighted that 21 % of recordings were inadequate and noted demographic-specific performance issues, emphasizing the discrepancy between lab and real-world performance.

**d. Geographic Distribution and Implementation Considerations**

Fig. 9 reveal notable geographic diversity: East Asia is the leading region (12 studies, 32.4 %), followed by Sub-Saharan Africa (8 studies, 21.6 %), North America (7 studies, 18.9 %), and South Asia (4 studies, 10.8 %). In Sub-Saharan Africa, an endemic region with a high RHD burden, Ethiopia showed research productivity with 6 studies [10,11,28,31,32], indicating a need for local technical investment. Despite commendable geographic distribution, the review identified a critical disconnect between technical development and implementation science. None of the 37 studies provided comprehensive cost-effectiveness analyses for their technologies, and only 3 studies (8.1 %) addressed practical deployment considerations such as device durability, battery requirements, or healthcare worker training. This gap is a substantial barrier to translating promising laboratory performance (median accuracy 93.7 %) into clinical impact in resource-constrained settings. Future research must prioritize prospective field validation in diverse settings and ensure integration with existing healthcare infrastructure to bridge this gap. Furthermore, the global research funding for Rheumatic Heart Disease (RHD) as of 2015 was only 0.1 % of the total funding (in contrast, HIV/AIDS, Tuberculosis, and Malaria covered 33.3 %, 18.6 %, and 18.6 % respectively) [62]. This funding level, which remains

**Table 8**
Characteristics of machine learning studies for detecting Combined Valvular Heart Diseases (VHD) & multi-condition detection.

| Ref | Year | Model | Preprocessing technique | Country | Sample size, data type, and access condition | Metrics | Limitation |
|---|---|---|---|---|---|---|---|
| [22] | 2022 | Hybrid CNN-LSTM | Downsampling, data augmentation (time stretch, time shift, noise addition, volume control), mono channel conversion, FFT (clipped to 350 Hz) | - | 1000 recordings, 5 classes) Source 2: PhysioNet/ CinC 2016 Challenge dataset | Five-class problem: Accuracy = 98.5 % F1-score = 98.501 % AUC = 0.9978 | Tested on limited datasets; lacks validation on diverse populations |
| [52] | 2023 | GoogleNet (Transfer Learning), Weighted-KNN | - CNN: Time series converted to time-frequency scalograms - KNN: Manual time/frequency domain feature extraction. | China | 1000 records (5 classes), PCG, Public | GoogleNet achieved 97.5 % accuracy | Focused on four heart valve diseases, excluding aortic regurgitation and tricuspid valve diseases. It did not compare its performance to expert cardiologists with the same samples. |
| [53] | 2023 | CNN (EfficientNet) + MLP fusion | ECG signals sampled at 500 Hz (5–10 s); median + Butterworth bandpass filtering (0.5–40 Hz); QRS/SD outlier removal | USA (New York City) | 617,338 ECG-TTE, private | AUROCs: AS 0.89 (internal), 0.86 (external); MR 0.88 (internal), 0.81 (external) | Limited to AS/MR only; low PPV (AS: 0.20); external validation within same health system; no analysis of asymptomatic cases; no murmur correlation |
| [54] | 2020 | (SRC, Nearest Neighbor Distances) | Chirplet Transform (CT) for time-frequency analysis, Butterworth bandpass filter, Shannon energy-based heart sound envelope extraction | India, Singapore, Taiwan, Japan | 800 PCG recordings (2400 cardiac PCG cycles) for HC, AS, MS, and MR classes; Publicly available GitHub database | Sensitivity for AS: 99.44 %, MS: 98.66 %, MR: 96.22 %, OA: 98.33 % | Limited dataset (800 PCG recordings), lower performance in MR classification, potential difficulty with overlapping signal characteristics |
| [55] | 2024 | CNNs and XGBoost | ECG signals sampled at 500 Hz for 10 s; input as 12 × 5000 matrix | Taiwan | 88,847 patients from two hospitals, retrospective ECG data, private | AUCs: AS >0.84, AR >0.80, PR >0.77, TR >0.83, MR >0.81 | Missing structured data for mitral stenosis; overestimation of prevalence; no murmur info; no clinical impact analysis; unclear performance in asymptomatic individuals |
| [55] | 2023 | 10-layer CNN, model stacking with Random Forest (RF) and XGBoost | PCG: Log-Mel spectrograms, STFT with Hann window, Mel-filter bank, SpecAugment, Mix-up. ECG: Cropping, resizing, horizontal shifting. | Germany /Japan | 1155 patients, PCG and 12-lead ECG, private | AUC: Aortic stenosis = 0.93, Mitral regurgitation = 0.80, Left ventricular dysfunction = 0.75 | Limited external validity: Single-center study without validation across diverse populations or settings. |
| [56] | 2023 | 1D Convolutional Neural Network (ValveNet) | ECG waveforms downsampled to 250 Hz; exclusion of paced/ poor-quality | USA | 77,163 patients, ECG, private | AUROC: AS: 0.88, AR: 0.77, MR: 0.83, Composite: 0.84; Sensitivity: 78 %, Specificity: 73 % | High false positives due to class imbalance; variation in ECG filtering; interobserver variability in echocardiogram reads; difficulty detecting AR; generalizability concerns |
| [57] | 2024 | EMAS AI algorithm (proprietary) | Applied noise cancellation, extracted 2-second segments with a 1-second slide, and filtered low-quality segments. | USA | 1029 participants; 4081 PCG recordings using EkoDUO/EkoCORE stethoscopes; ECHO as reference; private | Sensitivity: 39.3 %, Specificity: 82.3 % Best for aortic stenosis: 88.9 % | Low sensitivity, demographic bias (0 % detection in Black/ African American participants) and high recording rate inadequacy (21 %) and no core lab ECHO review and small sample sizes for severe VHD |
| [58] | 2020 | Time Growing Neural Network (TGNN), 3-layer perceptron | Downsampling to 2 KHz, antialiasing filter, spectral content calculation using forward/backward/mid-growing window | Iran | 15 pediatric patients (25 NM, 25 IM, 25 VSD, 10 ASD, 15 MR, 15 TR); private | Accuracy: 91.6 % ±3.9, Sensitivity: 88.4 % ±5.7, Avg. classification error: 9.89 % | Limited dataset size; needs more diverse and larger training data |
| [59] | 2024 | Two-layer LSTM + fully connected layer | Spike removal, downsampling (2205 Hz), cardiac cycle segmentation (modified Springer algorithm), MFCC extraction (Hanning window, 25 ms step), normalization (mean/SD) | - | 2124 patients (Tromsø Study); 8496 heart sound recordings (4 positions); retrospective, Private | AS Detection: Sensitivity 90.9 %, Specificity 94.5 %. AR/MR Detection: AUC 0.634 (AR), 0.549 (MR); improved to 0.766/ 0.677 with clinical variables. | No external validation; small VHD cases ($n$ = 51 AS, $n$ = 150 AR, $n$ = 292 MR); controlled recording environment; limited generalizability to asymptomatic cases. |
| [60] | 2021 | 1D CNN, and RNN (BiLSTM) | Signal duration standardization, Wavelet smoothing, and Z-score normalization | UAE | 1000 PCG samples (Normal and VHD), private | Accuracy: 99.32 % AUC: 0.998 Sensitivity: 98.30 % Specificity: 99.58 % | Not prospectively tested with new patients in a real clinical setting. Lacks evaluation against advanced architectures like attention mechanisms or transformers. |
| [61] | 2023 | DL - vision transformers (ViT) | Applied Z-score normalization, CWT for 2D TFR conversion, MFCC/LPCC feature extraction, pre-emphasis filtering, Hamming windowing, and 10-fold cross-validation. | South Korea & USA | 1000 audio files (200 per class) - normal, AS, MR, MS, MVP, Public | Accuracy: 99.90 % F1-score: 99.95 % | No prospective testing with new patients. Also no comparison with standard practices or experienced cardiologists' diagnoses, nor does it assess robustness against noise artifacts in real-world PCG recordings. |

markedly lower in low- and middle-income countries, must be substantially improved to reduce these structural inequities [63], (Fig. 10).

**e. Comparative Strengths of PCG vs ECG for early-stage RHD**:

Across the 37 studies, PCG and ECG offered distinctly different contributions to RHD detection. PCG models [11,12,29,36,37,40,45,51,54,57,60] performed best because rheumatic mitral and aortic lesions produce characteristic systolic or diastolic murmurs, giving PCG a direct acoustic signature of valve pathology. ECG-based approaches [32,35,39,41,43,44,47,49,51,53,55], by contrast, detected RHD only indirectly through later-stage structural consequences such as left atrial enlargement or ventricular strain caused by chronic valvular dysfunction, rather than the valve lesion itself. As a result, neither modality showed strong performance for truly early or subclinical RHD, where murmurs are faint and structural remodeling is minimal.

**f. Real World Feasibility: ECG/PCG Data Analysis Pipeline for Heart Disease Detection**

Regarding real-world feasibility, none of the 37 studies evaluated practical implementation factors such as device integration, connectivity requirements, workflow fit, or usability for frontline health workers, and only a few mentioned deployment considerations in passing [38,44,56]. Fig. 5 outlines the end-to-end workflow for ML-based RHD detection from raw ECG/PCG acquisition and preprocessing to feature extraction, model development, and performance evaluation while highlighting persistent gaps in multimodal integration and real-world clinical validation. Fig. 6 provides a practical overview of how ECG/PCG screening can be implemented in the field, with basic signal acquisition performed by frontline staff and automated machine-learning analysis producing an immediate diagnostic output (Figs. 7 and 9).

g. Future Research Directions and Opportunities

Our comprehensive analysis identified several high-priority areas for future research that would address current limitations and advance the field toward clinical implementation (Fig. 10).

1. **Standardized evaluation frameworks:** Creating standardized benchmark datasets with uniform preprocessing and evaluation metrics would facilitate meaningful comparisons between algorithms and boost progress. The PhysioNet/CinC Challenge model exemplifies this approach in heart sound classification.
2. **Prospective validation in endemic settings:** Future studies should focus on prospective, pragmatic trials, setting predefined performance thresholds and comparing results with clinical examination and echocardiography.
3. **Severity stratification:** Further studies should move beyond binary classification to automated staging of RHD severity. This stratification would enhance clinical utility of developed AI models pointing out patients requiring urgent intervention.
4. **Explainable AI techniques:** Developing interpretable models that highlight the signal features driving classification decisions would enhance clinician trust and potentially generate new insights about subtle RHD manifestations not currently recognized in clinical practice.

h. Strengths and Limitations of This Review

This scoping review has notable strengths, including a comprehensive search across five major databases from 2015–2025, offering a broad overview of RHD-associated valvular pathologies. However, it has limitations: excluded non-English publications, recent preprints and conference proceedings. Additionally, focusing on ECG and PCG-based approaches excludes promising work in simplified echocardiographic screening that could enhance RHD detection.

## 5. Conclusion

This comprehensive scoping review of 37 studies from 2015 to 2025 shows that machine learning applications using ECG and PCG signals present a transformative opportunity for accessible rheumatic heart disease screening, especially in resource-limited settings where traditional diagnostic methods fall short. The shift from classical algorithms to advanced deep learning models, notably convolutional neural networks achieving median accuracies of 97.75 %, demonstrates the technical feasibility of AI-driven auscultation as a practical alternative to conventional screening methods. Yet, this promise is clouded by significant methodological issues, such as small sample sizes, limited external validation, demographic biases, and a worrying gap between lab results and real-world readiness. This research matters beyond technical success, as it tackles a critical global health equity challenge: rheumatic heart disease mainly impacts low- and middle-income countries but remains often undiagnosed due to weak infrastructure. Although the geographic focus shows promising input from endemic regions like Sub-Saharan Africa and South Asia, most studies (73 %) relied on single-center data, and only 10.8 % included external validation. This highlights a major gap between model development and clinical implementation. The lack of cost-effectiveness and feasibility studies across all reviewed works emphasizes the need to connect technical advances with practical healthcare solutions, especially since conventional auscultation can miss up to 90 % of RHD cases in these areas. Moving forward, realizing AI's full potential in cardiac screening calls for a shift toward implementation science, standardized evaluation methods, and prospective validation in endemic regions. Future work should focus on creating benchmark datasets, conducting multi-center validation, and evaluating deployment factors like cost, healthcare worker training, and integration into existing clinical workflows. By overcoming these issues through rigorous trials and standard procedures, AI-based screening could ultimately provide timely, equitable cardiovascular diagnostics to underserved populations, potentially preventing thousands of disability-adjusted life years through early treatment and helping reach the World Heart Federation's aim of reducing RHD-related deaths by 25 % by 2025.

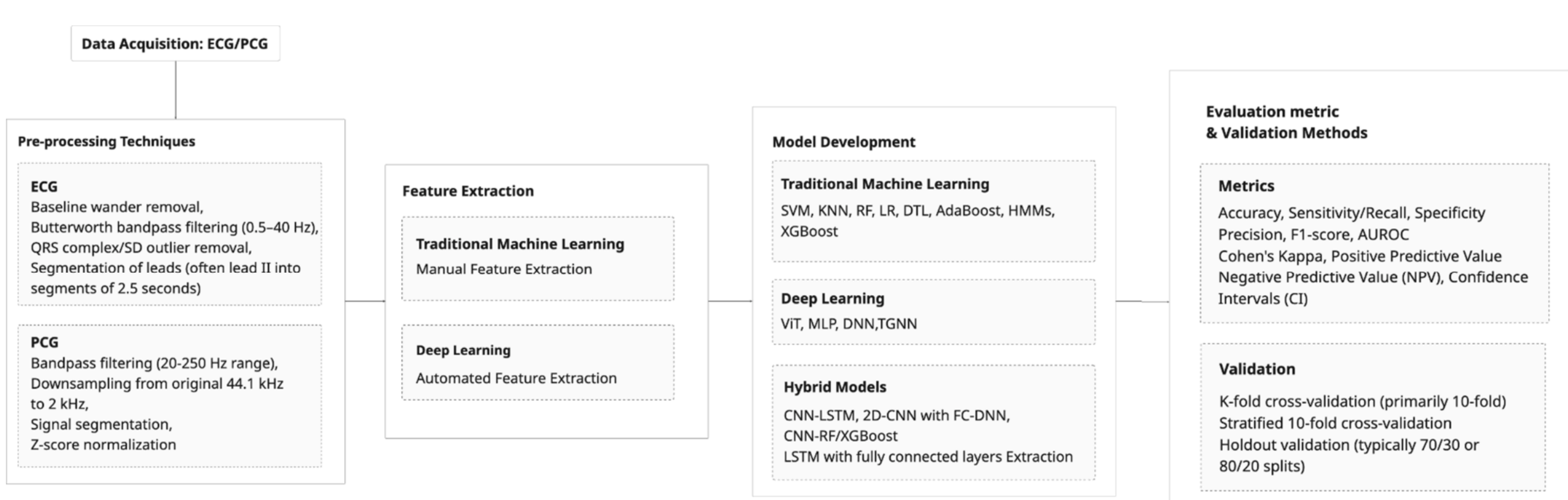

**Fig. 5.** ECG/PCG Data Analysis Pipeline for Heart Disease Detection - Pipeline workflow from signal acquisition through preprocessing to model development (traditional ML, deep learning, hybrid methods) and evaluation metrics.

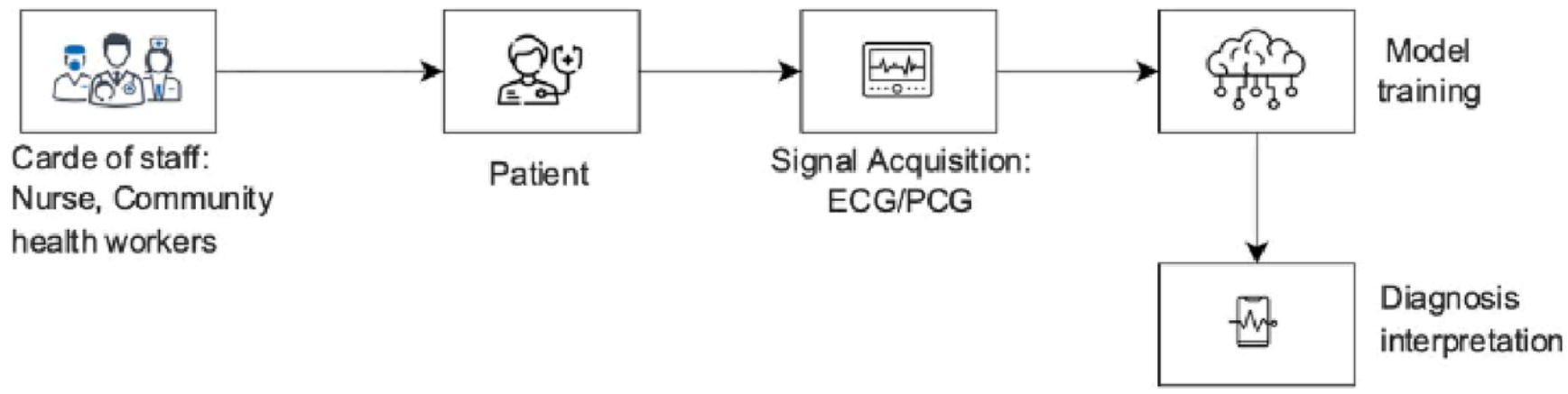

**Fig. 6.** Operational pipeline for usage for ECG/PCG screening using machine learning. Cadre-level staff perform basic signal acquisition, after which automated ML processing generates a diagnostic interpretation.

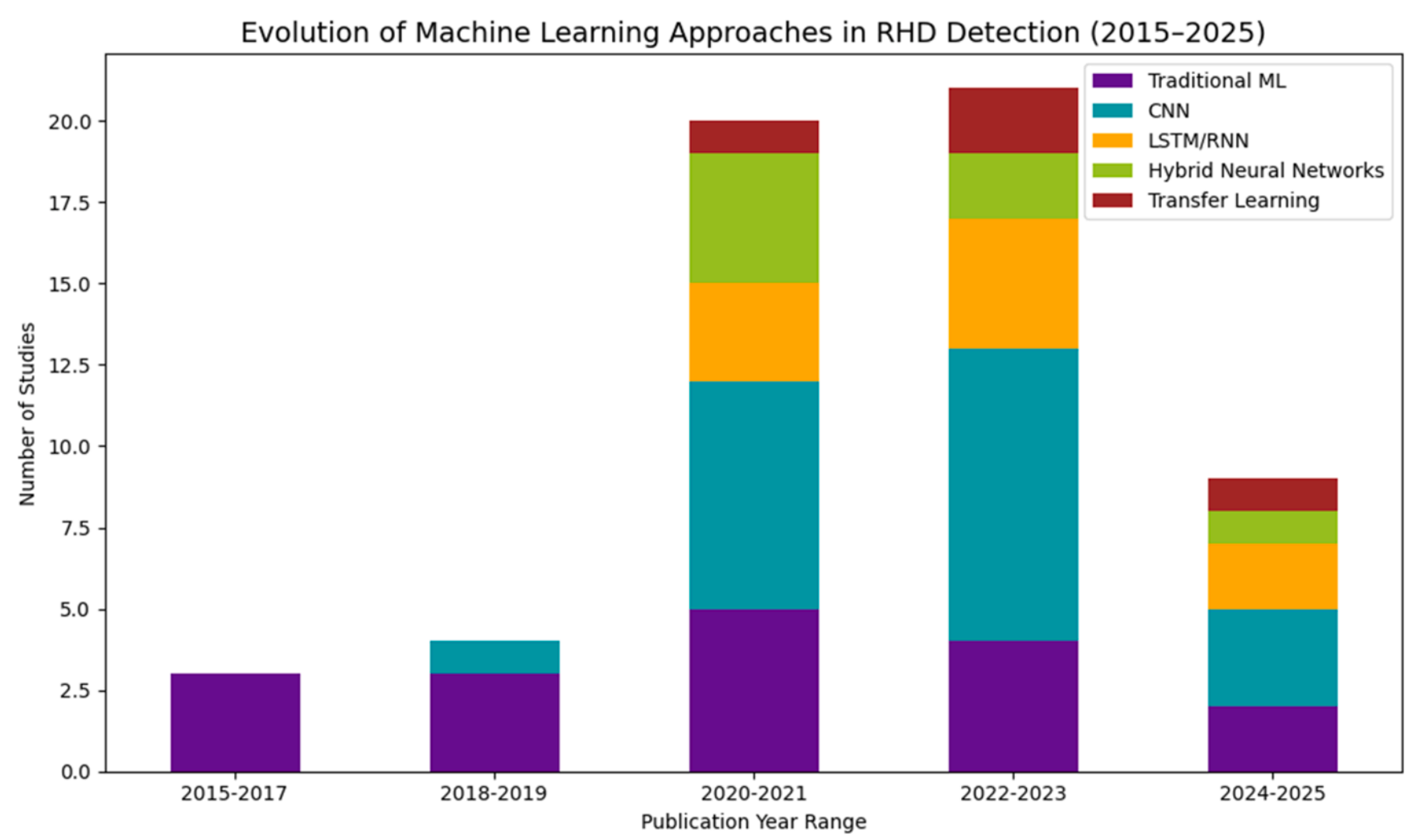

**Fig. 7.** Evolution of ML techniques for RHD detection (2015–2025). Research progressed from traditional ML methods (purple) to CNN-dominated approaches (teal) by 2020–2023, with emerging hybrid and transfer learning techniques. Peak activity occurred during 2020–2023. (Some studies used multiple techniques).

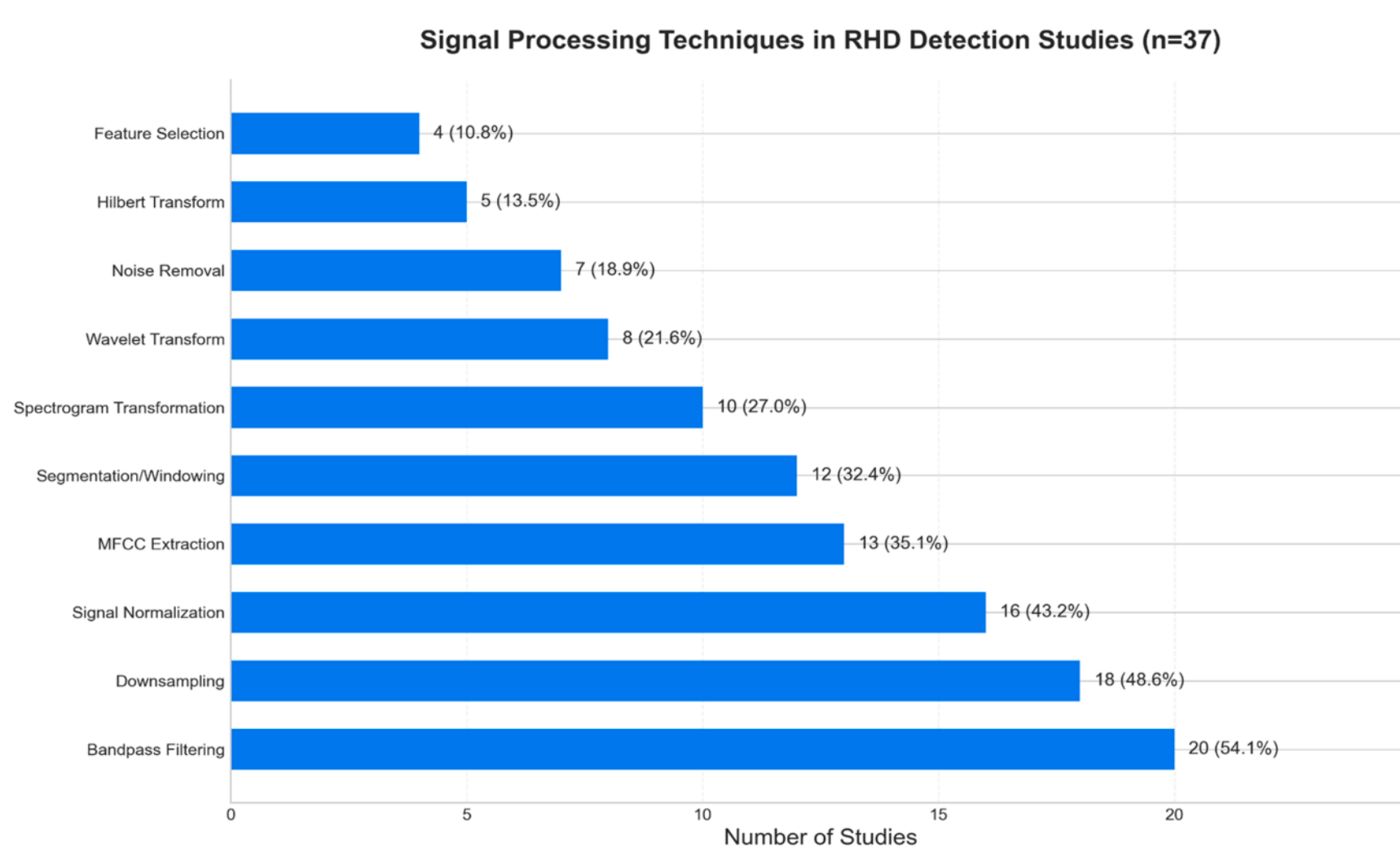

**Fig. 8.** Signal processing techniques in RHD detection studies ($n = 37$). Bandpass filtering was the most common preprocessing technique (54.1 %), followed by downsampling (48.6 %) and normalization (43.2 %). MFCC extraction and segmentation were prevalent feature engineering methods.

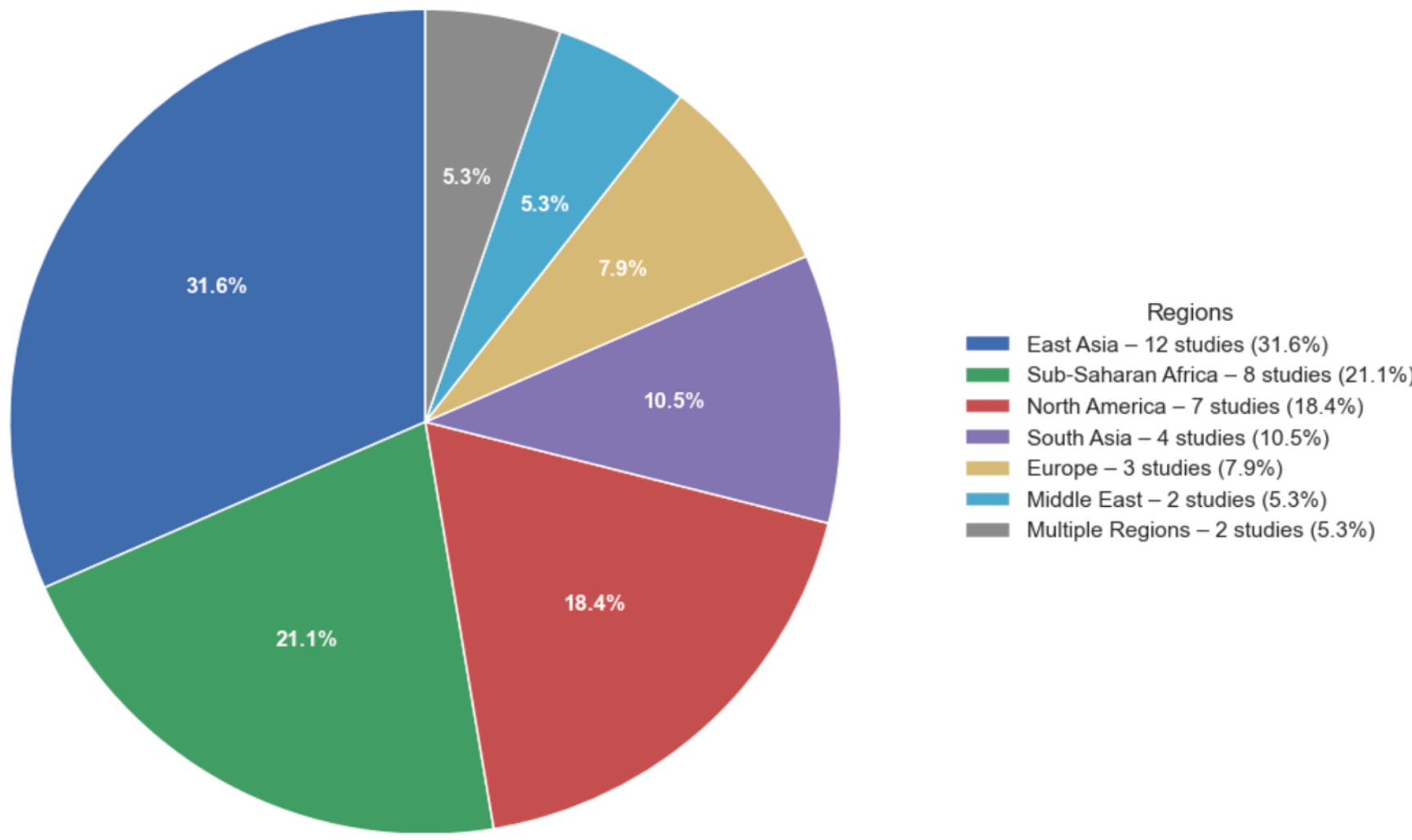

**Fig. 9.** Geographic distribution of machine learning for RHD Detection ($n = 37$).

| Disease | 2013 Deaths* | 2013 Funding in U.S. Dollars† |
|---|---|---|
| HIV–AIDS | 1,341,000 | 1,091,000,000 |
| Tuberculosis | 1,290,300 | 559,000,000 |
| Malaria | 854,600 | 533,000,000 |
| Rheumatic fever | 275,100 | 900,000 |

**Fig. 10.** Deaths from and Research and Development Funding for Rheumatic Fever and other Global Infectious Diseases. Data are from GBD 2013 Mortality and Causes of Death Collaborators. Despite causing over 275,000 deaths in 2013, Rheumatic fever received only a negligible fraction of R&D resources compared to diseases with higher mortality rates. The massive funding discrepancy highlights RHD's status as a Neglected Tropical Disease (NTD) or, more accurately, a neglected Poverty-Associated Disease.

## Ethical statement

Hereby, I, **Damilare Emmanuel Olatunji,** consciously assure that for the manuscript "**Machine Learning-Based Analysis of ECG and PCG Signals for Rheumatic Heart Disease Detection: A Scoping Review (2015–2025)**" the following is fulfilled:

1. The research was conducted in accordance with applicable laws and institutional guidelines at Carnegie Mellon University Africa, Rwanda Campus.
2. The research does not involve experiments with human subjects. Therefore, no privacy rights of human subjects were violated.
3. This material is the authors' original work and has not been published elsewhere before.
4. The paper is not currently being considered for publication elsewhere.
5. The paper reflects the authors' own research and analysis in a truthful and complete manner.
6. The paper properly credits the meaningful contributions of co-authors and co-researchers.
7. The results are appropriately placed in the context of prior and existing research.
8. All sources used are properly disclosed (correct citation).
9. All authors have been personally and actively involved in substantial work leading to the paper, and will take public responsibility for its content.

The violation of the Ethical Statement rules may result in severe consequences.

I agree with the above statements and declare that this submission follows the policies of [Journal Name] as outlined in the Guide for Authors and in the Ethical Statement.

**Damilare Emmanuel Olatunji** Carnegie Mellon University Africa College of Engineering Kigali, Rwanda dolatunj@andrew.cmu.edu 1st September 2025

## Declaration of interest

### *Financial*

The authors declare that they did not receive any funding or payments related to the publication of this study, nor do they anticipate receiving any in the future. Additionally, no financial support was provided by any organization with a vested interest in the outcome of this research. As mentioned in the manuscript, this study did not receive any specific grants from funding agencies in the public, commercial, or non-profit sectors.

### *Non-financial*

The authors declare that they have no non-financial interests that could benefit either themselves or their institution in terms of goods in kind. None of the authors hold unpaid advisory positions in government, non-governmental organizations, or commercial entities that would benefit from this research.

### *Affiliations*

All authors are affiliated with Carnegie Mellon University Africa, College of Engineering, located in Kigali, Rwanda. This affiliation does not present a conflict of interest, as the institution has no commercial or financial stake in the results of this systematic review.

### *Intellectual property*

The authors state that they and their organization hold no patents, trademarks, or intellectual property related to this work. There are no potential opportunities for exploiting the scientific advancements reported for the institution, the authors, or any research funders.

*Personal*

The authors declare no personal relationships (friends, family, or other close personal connections) with individuals who may potentially benefit financially or in other ways from this research.

*Ideology*

The authors have no political or religious beliefs, activism, or affiliations with advocacy or lobbying organizations that would create a conflict of interest in this work.

*Summary declaration*

The authors state that there are no competing interests, conflicts of interest, or potential biases related to this systematic scoping review on machine learning applications for the detection of rheumatic heart disease.

**Funding sources**

This research did not receive any specific grant from funding agencies in the public, commercial, or not-for-profit sectors.

**CRediT authorship contribution statement**

**Damilare Emmanuel Olatunji:** Writing – review & editing, Writing – original draft, Conceptualization. **Julius Dona Zannu:** Validation, Software. **Carine Pierrette Mukamakuza:** Project administration, Investigation, Funding acquisition. **Godbright Nixon Uiso:** Visualization, Validation. **Chol Buol:** Writing – original draft, Visualization. **John Bosco Thuo:** Project administration, Formal analysis. **Nchofon Tagha Ghogomu:** Writing – review & editing, Data curation. **Mona Mamoun Mubarak Aman:** Formal analysis, Data curation. **Evelyne Umubyeyi:** Methodology, Investigation.

**Declaration of competing interest**

None.

**Acknowledgment**

All co-authors (Julius Dona Zannu, Carine Pierrette Mukamakuza, Godbright Nixon Uiso, Chol Buol, John Bosco Thuo, Nchofon Tagha Ghogomu, Mona Mamoun Mubarak Aman, Evelyne Umubyeyi) have reviewed and agree with this ethical statement.

## Appendix

**Table I**
Systematic PubMed search strategy for identifying machine learning applications in rheumatic heart disease detection using electrocardiographic and phonocardiographic signals (2015–2025).

| Concept Block | Search Terms |
|---|---|
| Machine Learning | "machine learning"[MeSH] OR "machine learning"[tiab] OR "deep learning"[tiab] OR "artificial intelligence"[MeSH] OR "artificial intelligence"[tiab] OR "neural network*"[tiab] OR "convolutional neural network*"[tiab] OR "CNN"[tiab] OR "deep neural network*"[tiab] OR "DNN"[tiab] OR "recurrent neural network*"[tiab] OR "RNN"[tiab] OR "support vector machine*"[tiab] OR "SVM"[tiab] OR "random forest*"[tiab] OR "decision tree*"[tiab] OR "gradient boosting"[tiab] OR "XGBoost"[tiab] OR "feature extraction"[tiab] OR "computer-aided diagnosis"[tiab] OR "automated detection"[tiab] OR "algorithm*"[tiab] OR "signal processing"[tiab] OR "pattern recognition"[tiab] OR "computational intelligence"[tiab] |
| ECG and PCG Signals | "electrocardiogra*"[MeSH] OR "electrocardiogra*"[tiab] OR "ECG"[tiab] OR "EKG"[tiab] OR "phonocardiogra*"[MeSH] OR "phonocardiogra*"[tiab] OR "PCG"[tiab] OR "heart sound*"[tiab] OR "cardiac sound*"[tiab] OR "cardiac signal*"[tiab] OR "heart murmur*"[tiab] OR "cardiac murmur*"[tiab] OR "auscultation"[MeSH] OR "auscultation"[tiab] OR "cardiac electrical activity"[tiab] OR "cardiac acoustic*"[tiab] OR "biomedical signal*"[tiab] OR "cardiac monitoring"[tiab] |
| Rheumatic Heart Disease | "rheumatic heart disease"[MeSH] OR "rheumatic heart disease*"[tiab] OR "RHD"[tiab] OR "rheumatic fever"[MeSH] OR "rheumatic fever"[tiab] OR "rheumatic valv*"[tiab] OR "mitral stenosis"[tiab] OR "valvular heart disease*"[tiab] OR "valvular disease*"[tiab] OR "mitral regurgitation"[tiab] OR "aortic regurgitation"[tiab] OR "rheumatic carditis"[tiab] OR "rheumatic valvulitis"[tiab] OR "rheumatic mitral valve"[tiab] OR "rheumatic valve disease"[tiab] |
| Date Restriction | ("2015/01/01"[PDAT]: "2025/03/01"[PDAT]) |
| Combined Search | (Machine Learning) AND (ECG and PCG Signals) AND (Rheumatic Heart Disease) AND (Date Restriction) |